# Substrate engineering of inductors on SOI for improvement of Q-factor and application in LNA

ARUN BHASKAR[1,2], JUSTINE PHILIPPE[1], VANESSA AVRAMOVIC[1], FLAVIE BRAUD[1], JEAN-FRANÇOIS ROBILLARD[1], CEDRIC DURAND[2], DANIEL GLORIA[2], CHRISTOPHE GAQUIERE[1] AND EMMANUEL DUBOIS[1]

[1] Univ. Lille, CNRS, Centrale Lille, Univ. Polytechnique Hauts-de-France, Yncréa Hauts-de-France, UMR 8520 - IEMN, F-59000 Lille, France

[2] STMicroelectronics, Crolles 38926, France

**ABSTRACT** High Q-factor inductors are critical in designing high performance RF/microwave circuits on SOI technology. Substrate losses is a key limiting factor when designing inductors with high Q-factors. In this context, we report a substrate engineering method that enables improvement of quality factors of already fabricated inductors on SOI. A novel femtosecond laser milling process is utilized for the fabrication of locally suspended membranes of inductors with handler silicon completely etched. Such flexible membranes suspended freely on the BOX show up to 92 % improvement in Q-factor for single turn inductor. The improvement in Q-factor is reported on large sized inductors due to reduced parallel capacitance which allows enhanced operation of inductors at high frequencies. A compact model extraction methodology has been developed to model inductor membranes. These membranes have been utilized for the improvement of noise performance of LNA working in the 4.9 – 5.9 GHz range. A 0.1 dB improvement in noise figure has been reported by taking an existing design and suspending the input side inductors of the LNA circuit. The substrate engineering method reported in this work is not only applicable to inductors but also to active circuits, making it a powerful tool for enhancement of RF devices.

## I. INTRODUCTION

Integrated inductors on Silicon-on-Insulator (SOI) technology are critical components for the proper design of monolithic RF/microwave integrated circuits. The realization of high quality factor (Q-factor) integrated inductors is important for several RF design cases like input impedance matching for LNA [1], high quality LC resonators for voltage controlled oscillators (VCO) [2], filters [3], mixers [4], power amplifiers [5]. Substrate is one of the most important limiting factors in determining the maximum Q-factor. This is especially true of process technologies which use parallel stacking of all metal layers for lowering ohmic resistance of inductor [6], [7]. In these technologies, small dielectric spacing between the substrate and inductor metal result in higher substrate coupling. This limits the maximum achievable Q-factor values on SOI technology thereby limiting overall circuit performance. A good comparison of different substrates is reported in [8], [9]. Superior performance of insulating quartz substrate over state-of-the-art High-Resistivity (HR) and Trap-Rich (TR) SOI substrates has been demonstrated in [8]. Thus, optimizing the substrate is a possible pathway for better RF circuit operation. In previous studies, improvement of electrical performance of RF devices has been reported by complete removal of handler substrate on SOI wafer and transfer onto thin flexible substrate [10]–[15]. While this approach has been shown to work efficiently, some of the drawbacks are higher number of steps in fabrication, possible RF losses in bonding material, mechanical weakening and overall cost of treatment.

A local handler substrate removal method enables suspension of RF circuits freely on the BOX and substrate related losses can be brought to a minimum. In this context, we have developed the Femtosecond Laser Assisted Micromachining and Etch (FLAME) process to fabricate membranes of RF circuits [16], [17]. Substrate removal has already been studied in numerous works as a means of improving Q-factor [18]–[22]. The novelty of the FLAME process is that it is performed after fabrication of the die is complete. It can be applied on any SOI process technology that is offered by any foundry making it highly versatile. Additionally, high material removal rates are possible making the process time-efficient. For instance, in our work on RF switches, volume removal rate of 8.5 x $10^6$ $\mu m^3 s^{-1}$ was reported [17]. In the following, this paper is hierarchically organized from process description to device and circuit characterization. First, the FLAME process is briefly described in section II which outlines the fabrication method of membranes of inductors. RF characterization of inductor membranes is subsequently described in section III to benchmark them against inductors on state-of-the-art SOI technology. Compact modelling methodology is presented in section IV which describes the process of obtaining lumped element models of suspended inductors. Finally, the application of inductor membranes to improve LNA performance is discussed in section V.

## II. FLAME PROCESS

Femtosecond laser material removal is the main process step in the fabrication of membranes using FLAME process (Fig. 1). According to this approach, a focused gaussian beam is raster scanned over the area where silicon needs to be removed as illustrated in Fig. 1b. The absorption of laser radiation in the material causes ablation of silicon. The boundaries of ablation are set by appropriately defining the area of scanning. The beam is scanned multiple times over the handler silicon until desired depth of ablation is reached. After ablation is complete, the remaining thickness of silicon is typically few tens of microns. To remove this final portion of silicon, xenon difluoride ($XeF_2$) dry etching technique is employed. $XeF_2$ is an isotropic etchant of silicon which has a high selectivity of etch of $Si:SiO_2$ of 1000:1 [23]. By using an etch mask to protect the unablated regions, membranes are obtained in the ablated areas after the completion of $XeF_2$ etch step leaving rest of the die intact. The process has been described in detail in previous works where mm sized membranes have been demonstrated with material removal rates of up to $8.5\times10^6$ $\mu m^3$ $s^{-1}$ [16], [17]. At the end of the process, silicon is removed completely under the inductor area (ABCD in Fig. 1a) and retained under the bond pads. The thickness of silicon under the bond pad is reduced to ~40-50 µm. Since most of the field lines are concentrated over this thickness, we can assume that the pad conditions do not change appreciably after FLAME process.

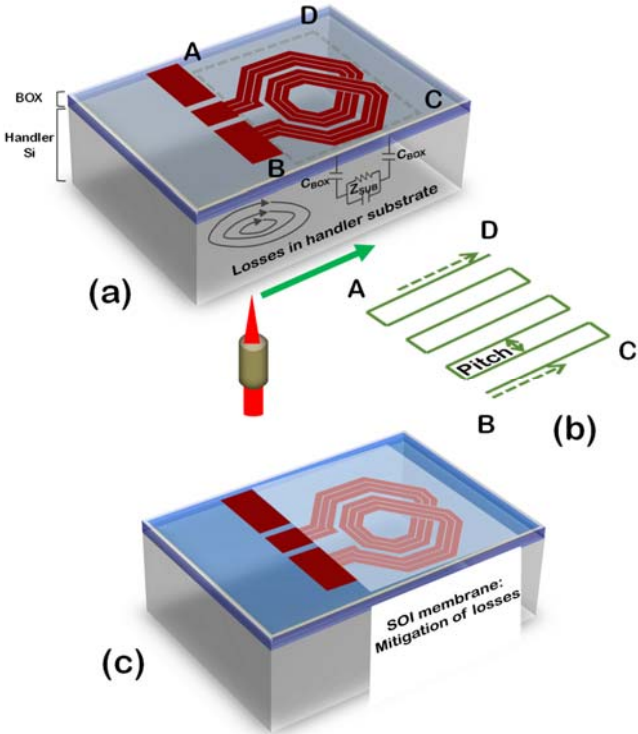

Fig. 1. FLAME process for fabrication of membranes on SOI wafers (a) Outlined area ABCD represents the region of substrate coupling (b) Laser trajectory on the handler silicon side to ablate area ABCD (c) Membrane fabricated using FLAME process

## III. RF CHARACTERIZATION OF MEMBRANES OF INDUCTORS

The inductors are fabricated using the 0.13 µm SOI-CMOS technology by STMicroelectronics (H9SOIFEM). The inductors use stacking of 4 parallel metal layers M1 – M4 (3 thin Al and 1 thick Cu capped with Al) with an effective metal thickness of 9.2 µm to obtain low self-resistance of the inductor. The spacing from M1 to $Si/SiO_2$ interface is 1.2 µm. Two types of SOI substrates are characterized before and after substrate removal: 1) High Resistivity SOI (HR-SOI) 2) Trap Rich SOI (TR-SOI). Both substrate types have a high resistivity of handler silicon (1-100 kΩ.cm). HR-SOI has more losses because of the parasitic surface conduction layer formed close to the back $Si-SiO_2$ interface. In TR-SOI, a layer of polysilicon is introduced at the interface to mitigate this effect in such a way that losses are considerably reduced [8], [24], [25]. The discussed results for membranes are also applicable to newer SOI wafer technologies like those involving PN doped implants to combat the interface parasitic conduction layer [26], [27].

S-parameters characterization is performed over a frequency range of 100 MHz – 110 GHz with a low IF bandwidth of 50 Hz to improve accuracy. Since a 1-port setup is used, and the parasitics are not negligible for the chosen inductance values, pad de-embedding is necessary. An open pad is used to determine the pad admittance ($Y_{pad}$) which is later used for de-embedding. The admittance of the measured inductor ($Y_{meas}$) contains contributions both by the pad and the inductor itself. The impedance of the inductor is simply calculated as:

$$Z_{ind} = (Y_{meas} - Y_{pad})^{-1} = R_{eff} + j.X_{eff} \qquad (1)$$

where $R_{eff}$ and $X_{eff}$ stand for the inductance effective resistance and reactance, respectively. The Q-factor of the integrated inductance is calculated as:

$$Q_{ind} = \frac{X_{eff}}{R_{eff}} \qquad (2)$$

The effective resistance ($R_{eff}$) is a frequency variable resistance which depends on losses in interconnect metal layers and handler substrate [28]. The reactance $X_{eff}$ is also frequency dependent but remains almost constant and mainly reflects an inductive behavior up to self-resonant frequency. It is well established that beyond self-resonance, the parasitic capacitance associated to the inductance turns of the wound metal lines dominates at high frequency. To evaluate the impact of the FLAME process, a single turn inductor of inductance ~0.85 nH is characterized before and after membrane suspension. The Q-factor curves are shown in Fig. 2. The peak Q-factor value increases from 26.5 to 51 for HR-SOI (92% improvement) and from 27.9 to 49.7 for TR-SOI substrate (78% improvement). After applying the FLAME process, it can also be observed that the Q-factor curves excellently match indicating that the handler substrate is completely removed under the inductor area without any leftover residues. It is worth noting that, in another work, for the same inductance value, Q-factor optimization was obtained by stacking two additional copper metal layers leading to a peak Q-factor of 34 at a frequency of 4.5 GHz [6]. This comparison shows that Q improvement is much higher when the substrate is removed, indicating that substrate removal approach is a powerful method of inductor optimization. The frequency of peak

Q-factor is also pushed to a higher value of ~13 GHz making it suitable for high frequency applications.

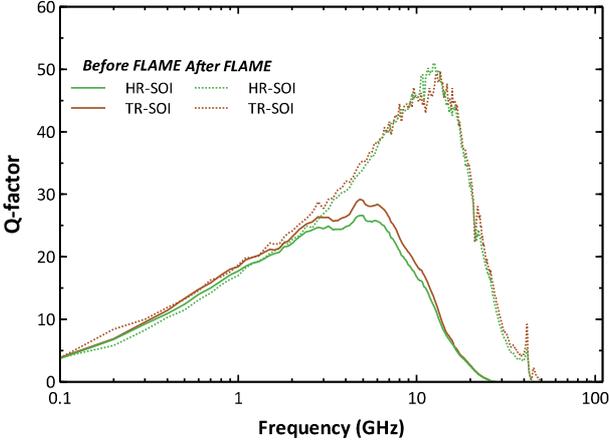

Fig. 2. Q-factor comparison before and after FLAME process for single turn inductor of inductance 0.85 nH originally integrated on HR-SOI and TR-SOI substrates.

While single turn inductors have been reported to get a picture of Q-factor improvement over previously reported work, multi-turn inductors are more commonly used in RF design. Hence, 2-turn inductors are also characterized and analyzed in more details in this work. The inductor dimensions are presented in Fig. 3 while the inductance value and inner diameter for each inductor are listed in Table I.

Contact resistance is an important consideration in the measurements as the self-resistance values are very small (< 0.5 Ω). Also, contact resistance changes depending on how well the probes are contacted to the bond pads. This typically changes for each measurement and hence estimation of Q-factor becomes challenging because of large dispersion of measured values for the same inductor. In order to remove the contribution of contact resistance from the measurement, EM simulation is performed and the self-resistance of the inductor is extracted at 100 MHz. Any measured resistance which is in excess of this value is counted as the contact resistance and subtracted from the measured value. The dispersion is efficiently reduced after this correction.

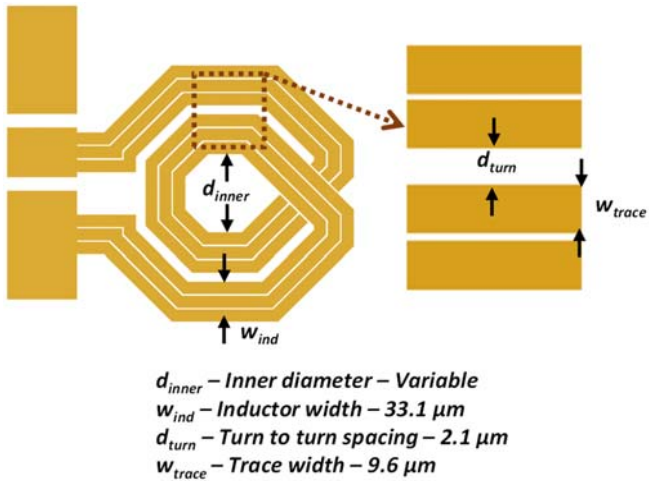

$d_{inner}$ – Inner diameter – Variable
$w_{ind}$ – Inductor width – 33.1 μm
$d_{turn}$ – Turn to turn spacing – 2.1 μm
$w_{trace}$ – Trace width – 9.6 μm

Fig. 3. Inductor layout depicting different dimensions

TABLE I: Details of 2 turn inductors used in the study.

| Inductor | Inner Diameter (μm) | Inductance (nH) |
|---|---|---|
| 2T-S | 80 | 0.6 |
| 2T-M | 240 | 1.8 |
| 2T-L | 400 | 3.3 |

The measured parameters for 2 turn inductors are shown in Fig. 4. Prior to substrate processing, it can be seen that for 2T-M and 2T-L, TR-SOI has superior Q-factor as compared to HR-SOI. For 2T-S, the difference is negligible indicating that substrate becomes more important as the size of the inductor increases. The inductor Q-factor for HR-SOI and TR-SOI substrate after FLAME process nearly match like in the case of single turn inductor with marginally higher values for HR-SOI as compared to TR-SOI. There are two possible reasons for the small observed differences. The first reason is that open pad de-embedding slightly overestimates the Q-factor because of the changed pad configuration as explained in section II. The second source is the variation in etched areas obtained from the FLAME process. This can be validated by looking at the boundaries of the membrane shown in Fig. 5. These are dual light microscope (DLM) images taken by passing backlight from the bottom side of the membrane and taking the image from the top side which has dim lighting. The area of membrane appears brightly in the image and this can be used to determine where the substrate is removed and where it is retained. For both substrate types, a very small part of the substrate is present under the inductor metal. This unetched area is marginally higher for TR-SOI as compared to HR-SOI. However, this effect is expected to be much smaller than the overestimation from pad de-embedding. The overall result is a 4-5% higher estimation of Q-factor in for HR-SOI as compared to TR-SOI. Taking TR-SOI as the reference, the increase in Q-factor is 58%, 47% and 35% for inductors 2T-S, 2T-M and 2T-L, respectively. The Q-factor improvement scales down with increasing size of inductors. Higher Q-factor after FLAME process can be attributed to the reduction of both parallel capacitance and effective resistance ($R_{eff}$) resulting from the suppression of losses in the substrate. This is further explained in section IV. Another interesting observation is that, at low frequencies, the Q-factor curves after FLAME process follow the curves before substrate removal. At these frequencies, there is no difference between HR-SOI and TR-SOI both before and after FLAME. Thus at lower frequencies, the substrate losses do not play a significant role in the performance of the inductor.

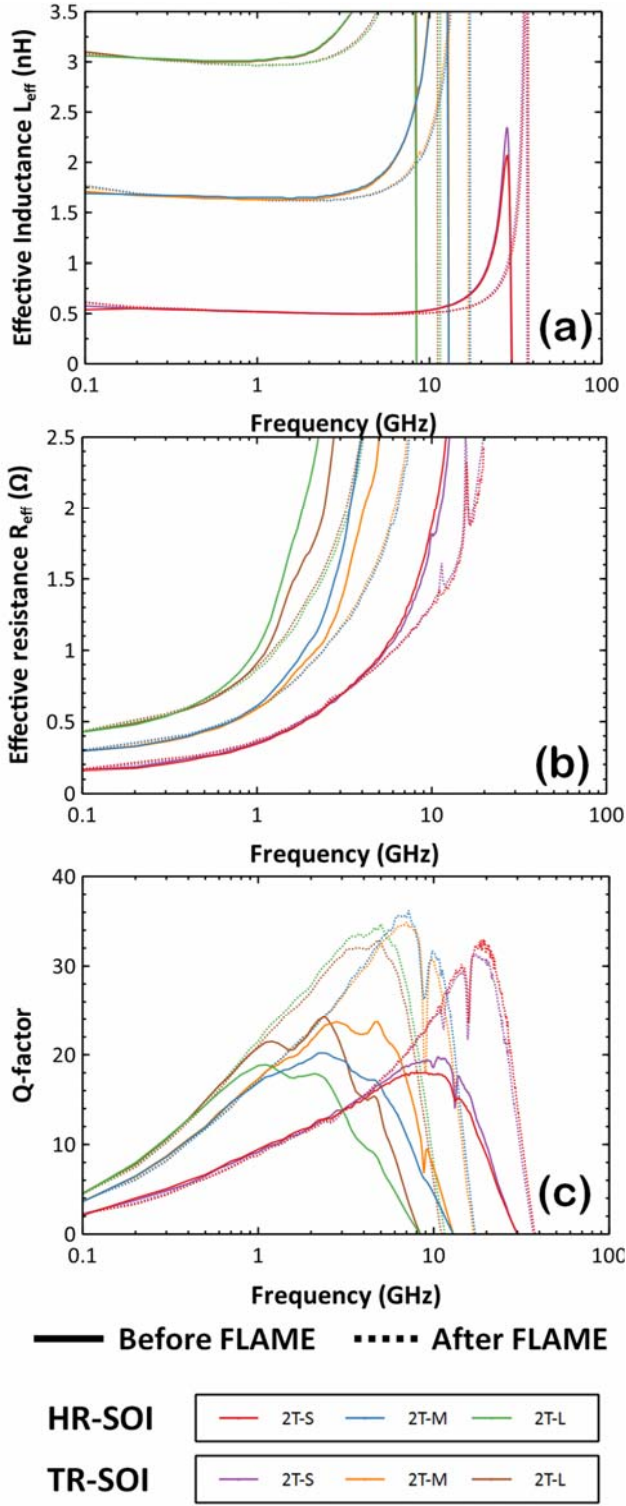

Fig. 4. Parameters for 2-turn inductors extracted from RF characterization (a) Effective inductance ($L_{eff}$) (b) Effective resistance ($R_{eff}$) (c) Q-factor ($Q_{ind}$)

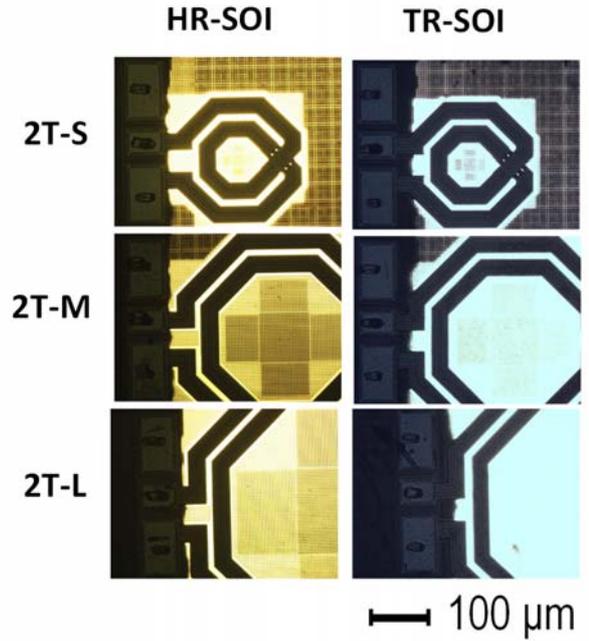

Fig. 5. DLM images depicting the boundaries of the membrane in the vicinity of the RF pad

## IV. COMPACT MODELLING OF MEMBRANES OF INDUCTORS

Small signal measurements discussed so far have clearly indicated the benefits of substrate removal for Q-factor improvement even in low loss substrates like HR-SOI and TR-SOI. For circuit design, it is beneficial to have a compact modelling of suspended inductors. For lossy substrates, pi-model is commonly used to describe the substrate loss mechanisms [29]–[31]. Based on specific design cases, in other works, the pi-model has been extended to accommodate skin and current crowding effects [32], [33]. Using this type of compact model, the behaviour of inductors in lossy substrates has been well modelled, revealing that the displacement and eddy currents flowing in the substrate limit the Q-factor of inductor.

These models do not work well for inductors with lossless substrates. A comprehensive study of inductors on nearly lossless Silicon-on-Sapphire substrates has been reported by Kuhn et al. [28]. This approach is taken as the basis for modelling of inductors on SOI membranes. When substrate losses are mitigated, inductor metal losses become the dominant mechanism. The inductor can be represented by a simple three-elements model as shown in Fig. 6a. In this model, $L_s$ represents the inductance and $C_p$ represents the capacitive element responsible for self-resonance. $C_p$ is the capacitance arising from turn to turn coupling in multi-turn inductors. These two elements are here considered frequency-independent. The self-resistance $R_s$ is frequency dependent to capture the two unavoidable loss mechanisms of current crowding and skin effect which are explained in [34] and [35].

For the 3-element model, an expression of the frequency dependent series resistance is to be calculated in order to model loss mechanisms. To establish a frequency independent model, Kuhn et al. proposed the 6-elements schematic shown in Fig. 6b which consists of a mutual inductance with resistors modelling the different frequency dependent metal losses. On this basis, a circuit transformation of the 6-elements model proposed in [36] is used to obtain a modified version comprising 2 inductors, 1 capacitor and 3 resistors. Specifically, the proposed model in Fig. 6c comprises of the first Cauer-form network (2nd order) of the outlined part in Fig. 6b. This is similar to the modelling methodology proposed by Ooi et al. in [37]. After obtaining the model elements for the membrane, the model for inductor on HR-SOI/TR-SOI substrate can be deduced using single-pi model by adding 2 additional elements $C_{par}$ and $R_{par}$ to the model as shown in Fig. 6d.

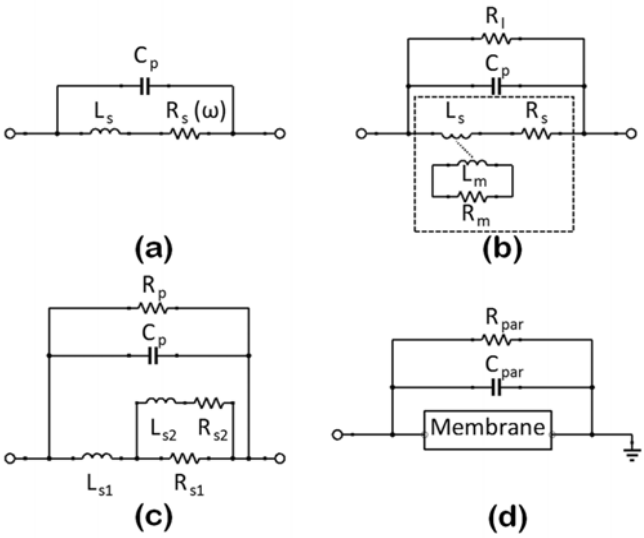

Fig. 6. Inductor models of low loss substrates (a) 3-elements frequency dependent model (b) 6-elements model with mutual inductance (c) Modified 6-elements model for inductor membranes (d) Inductor model with original substrate for single-ended excitation.

The procedure for determining the elements of the model is outlined as following:
a) Compute the pad-deembedded complex impedance of the inductor from the S-parameters characterization to obtain:
$$Z_{ind} = R_{eff} + j \cdot X_{eff} \quad (3)$$
b) $L_{s1}$ is calculated from the reactance at a low frequency here taken at 1 GHz:
$$L_{s1} = \left.\frac{X_{eff}}{\omega}\right|_{1GHz}. \quad (4)$$
c) With the increase in frequency, $X_{eff}$ drops to zero as the inductor parasitic capacitance compensates the inductive part. This corresponds to the frequency of self-resonance $\omega_r$ that is used to calculate $C_p$ as:
$$C_p = \frac{1}{L_{s1}\omega_r^2}. \quad (5)$$
d) For the resistive part of the inductor impedance, the dominant elements in the model are $Rs_1$ and $Rs_2$. Since they appear in parallel, the following condition is set at low frequency:
$$\left.R_{eff}\right|_{100MHz} = R_{s1}||R_{s2} \quad (6)$$

e) Finally, the parameters $R_p, R_{s1}, L_{s2}$ are varied to obtain the best fit. $R_{s2}$ is not an independent variable because of the constraint imposed in step d). Additionally, $L_{s1}$ is also adjusted for each value of $L_{s2}$ used during the fitting process. This is because $R_{s2} - L_{s2}$ branch which models frequency dependent resistance also adds an inductive component of reactance which increases the total inductance. $L_{s1}$ is changed such that the value of inductance calculated from the model is equal to $\left.\frac{X_{eff}}{\omega}\right|_{1GHz}$.

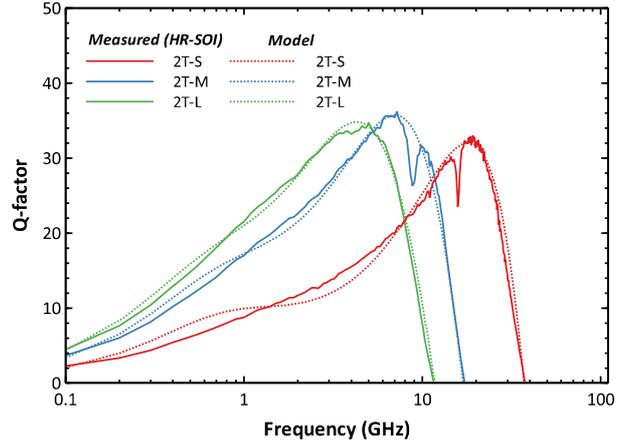

Fig. 7. Comparison of measured vs. modelled inductor Q-factors after FLAME process

The results of modelling are depicted in Fig. 7. The model elements extracted for the modified 6-elements model is tabulated in Table II. Both the peak value of Q-factor and the frequency at which it occurs are captured well using the modified 6-elements model. There is a small deviation between measured and modelled Q-factor at frequencies less than the peak Q-factor. This deviation comes from the fact that a 2nd order model cannot precisely capture the self-resistance behaviour of the inductor. By using a higher order model, precision can be potentially improved. However, the modelling process becomes more cumbersome which may not be desirable.

TABLE II: Modified 6-element model parameters for 2-turn inductors (for suspended inductors after FLAME process)

| Inductor | 2T-S | 2T-M | 2T-L |
|---|---|---|---|
| $L_{eff}$ @ 1 GHz (nH) | 0.61 | 1.81 | 3.25 |
| $R_{eff}$ @ 100 MHz (Ω) | 0.17 | 0.3 | 0.43 |
| $R_{s1}$ (Ω) | 1.1 | 1.2 | 1.4 |
| $R_{s2}$ (Ω) | 0.2 | 0.4 | 0.62 |
| $L_{s1}$ (nH) | 0.49 | 1.55 | 2.9 |
| $L_{s2}$ (nH) | 0.1 | 0.17 | 0.28 |
| $R_p$ (kΩ) | 13 | 11 | 10 |
| $C_p$ (fF) | 37.5 | 57 | 63.5 |

Some important observations can be made by looking at the values of the model parameters in Table II. Both $R_{s1}$ and $R_{s2}$ increase with the increase in size (diameter) of the inductor. This is because the resistance depends on the length of the conductor which increases with inductor diameter. $L_{s1}$ and $L_{s2}$ increase similarly due to inductance value dependence on the length of conductor. It is to be noted that higher $L_{s2}$ means that it blocks current flowing through $R_{s2}$ starting from lower frequencies. Referring to the effective resistance curve in Fig. 4b, it can be observed that it initially increases slowly and rises sharply beyond a certain frequency. This frequency is higher for inductors of smaller size. In the model, the frequency of transition is captured by $L_{s2}$. Higher $L_{s2}$ means that the knee point occurs at smaller frequencies. $R_p$ is used for fine adjustment of peak Q-factor and corresponding frequency. Finally, $C_p$ models the frequency of self-resonance where the inductor value and the Q-factor drop to zero. Smaller inductors means smaller coupling between turns. Hence, for small size inductors, lower $C_p$ shifts the self-resonant frequency to higher values. For the smallest inductor, the self-resonant frequency is ~37.5 GHz.

capacitance is in the pF range (RF short) and can be ignored at GHz frequencies. The value of $C_{par}$ is determined by fixing arbitrary $R_{par}$ and finding the value of $C_{par}$ which gives the same self-resonant frequency as seen in the measurement. Next, $R_{par}$ is adjusted such that best fit of Q-factor is obtained with respect to the measured data. The measured and modelled Q-factors with substrate are shown in Fig. 8. The computed values of $R_{par}$ and $C_{par}$ for each case is listed in Table III. It can be seen that the model gives slightly pessimistic values of Q-factor as compared to measured values especially around the region of the peak. This can be attributed to the fact that the model overestimates effective resistance in this region. As a result the modelled peak Q-factor is 8-14% lesser than the measured peak Q-factor. The value of $C_{par}$ does not change for the two substrates while $R_{par}$ is lower for HR-SOI as compared to TR-SOI as expected due to parasitic surface conduction effect in HR-SOI. With the extracted $R_{par}$ values in the kΩ range, the losses in both substrate types are not negligible. This is consistent with observation made by Liu et al. where a clear difference in Q-factor was observed between HR-SOI, TR-SOI and quartz substrate types [8].

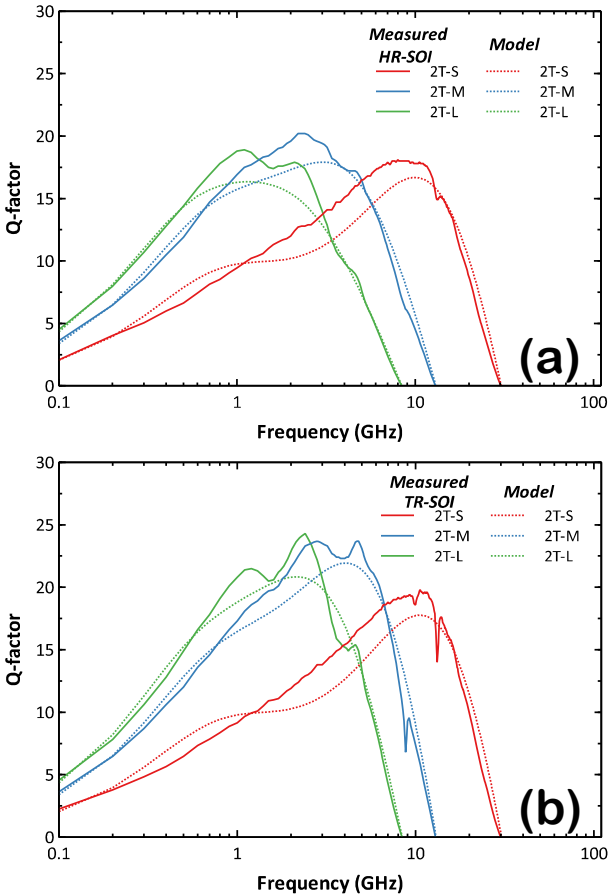

Fig. 8. Comparison of measured vs. modelled inductor Q-factors for HR-SOI and TR-SOI substrates

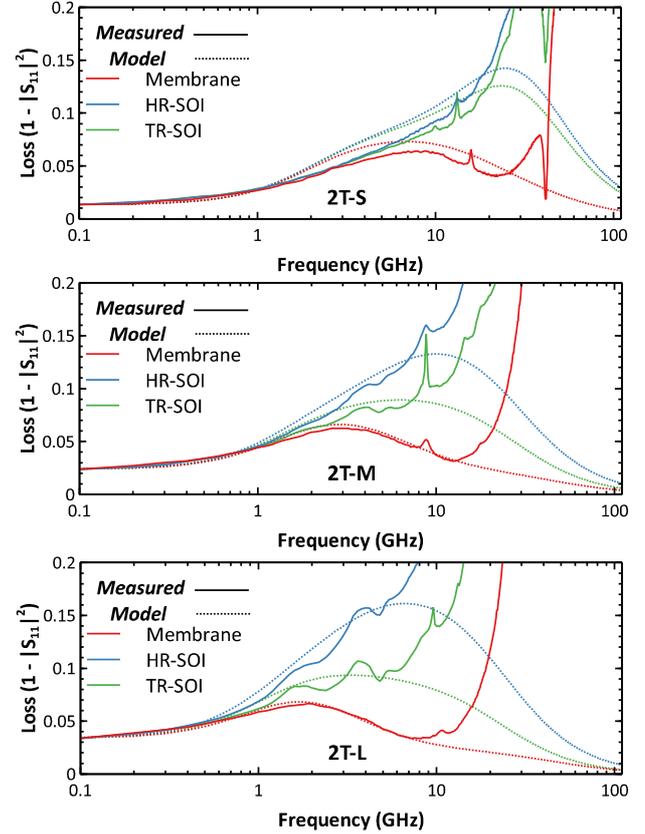

Fig. 9. Comparison of inductor losses before and after substrate removal

After having obtained the compact models for the membranes, the effect of substrate can be captured by addition of two elements $C_{par}$ and $R_{par}$ as seen in Fig. 6d. When substrate is present, the parallel capacitance due to turn-to-turn coupling changes because of the modification of the dielectric configuration. Additionally, substrate coupling occurs through the oxide capacitance and losses occur in the substrate. Both of these phenomena combined can be captured by two elements $C_{par}$ and $R_{par}$ in a single ended configuration. The oxide

In order to compare the inductor losses, the plots of $1 - |S_{11}|^2$ are shown in Fig. 9. The difference in losses with substrate depends on the size of the inductor. With increase in size, the difference in losses between HR-SOI and TR-SOI becomes larger. The same point is also reflected in the fact that difference in peak Q-factor increases for larger inductors for the two substrate types. After substrate removal, the losses are reduced because of replacement of handler silicon by air. In the model, a reduction in losses is seen at higher frequencies. This is not valid because the compact model captures inductor behaviour only until the self-resonant frequency is reached.

TABLE III: Modified 6-element model parameters for 2-turn inductors (for suspended inductors after FLAME process)

| Inductor | 2T-S | 2T-M | 2T-L |
|---|---|---|---|
| HR-SOI | | | |
| $C_{par}$ (fF) | 19 | 40 | 61 |
| $R_{par}$ (Ω) | 1785 | 1820 | 1360 |
| TR-SOI | | | |
| $C_{par}$ (fF) | 19 | 40 | 61 |
| $R_{par}$ (Ω) | 2200 | 3800 | 3480 |

## V. APPLICATION OF INDUCTOR MEMBRANES IN LNA CIRCUIT

For demonstration of practical applicability of the inductor membranes, a Wi-Fi LNA circuit is chosen as the test vehicle. The LNA that is studied has an operating range of 4.9 – 5.9 GHz and common source configuration with inductive source degeneration as shown in Fig. 10a and described in [38]. The common source LNA gives a good noise performance compared to other implementations. By improving the quality factor of the input inductors, the noise performance of the LNA can further be improved. This has been established in the noise analysis for common source LNA circuits by Shaeffer et al. [39]. The input stage of common source LNA as analyzed by Shaeffer et al. shows that the losses in the matching gate and source degeneration inductors contributes to the overall noise figure of the LNA. The same observation has been made in other works as well [1], [40].

The LNA dies used in this work are fabricated on a 750 μm thick substrate. They are grinded to reduce the starting thickness to ~250 μm before applying the FLAME process. A two-step laser milling strategy is used to balance the removal rate and milling quality as described in [17]. The laser parameters used for processing along with the attained removal rates are shown in Table IV. The inductors $L_{in}$ and $L_{deg}$ are suspended to evaluate the overall effect of improved quality factor of input inductors on LNA performance. The DLM images of the suspended inductors are shown in Fig. 10b. Membrane suspension is performed only on the HR-SOI substrate as it has been already demonstrated that after substrate removal, the performance of the two substrate types converge.

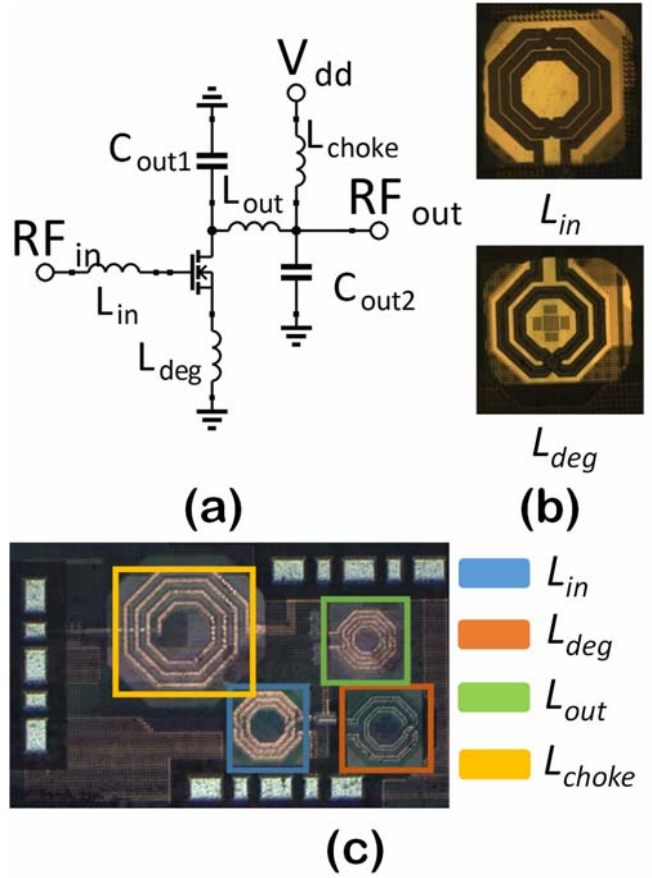

Fig. 10. (a) LNA schematic (b) DLM images of suspended inductors of LNA (c) Optical microscope image of the LNA showing different inductors

TABLE IV: FLAME process parameters used for fabrication of membranes

| Parameter | Fast step | Slow step |
|---|---|---|
| Scanner speed (mm s$^{-1}$) | 20 | 5 |
| Pulse repetition freq (kHz) | 30 | 2 |
| Laser power (W) | 1.006 | 0.013 |
| Number of passes | 4 | 180 |
| Fluence (J cm$^{-2}$) | 32.1 | 6.4 |
| Removal rate (x $10^6$ μm$^3$ s$^{-1}$) | 4.75 | 0.02 |

Small signal S-parameters and noise figure have been measured for the LNA on the same bench with a source impedance of 50 Ω. Before FLAME process, three dies have been characterized for each type of substrate, HR-SOI and TR-SOI. The average measured noise figure and gain are plotted in Fig. 11. The noise measurement shows very little data scatter across the 3 devices for both HR-SOI and TR-SOI substrates. The minimum NF is obtained at ~6 GHz. with TR-SOI showing ~0.09 dB lower average noise figure as compared to HR-SOI substrate in frequency range 4.9 – 5.9 GHz. All reported data henceforth is the average in the same range unless mentioned otherwise.

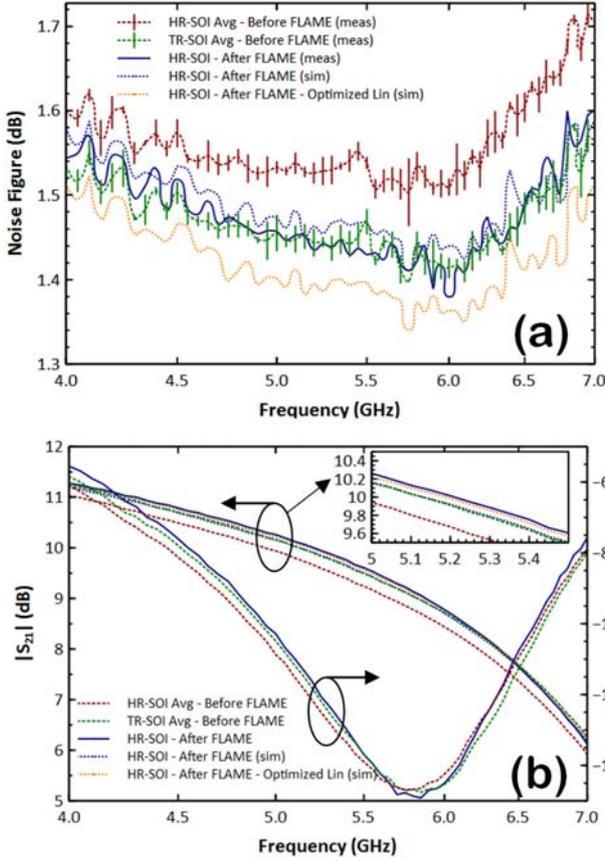

Fig. 11. Small signal parameters before and after FLAME process
(a) Noise Figure (b) |S$_{21}$| and |S$_{11}$| with inset showing |S$_{21}$| in the frequency range 5-5.5 GHz.

Since the difference in noise figures of HR-SOI and TR-SOI substrate is very small, care has to be taken to make sure that there are no errors due to calibration. A reference device is measured each time after calibration to ensure that all measurements are accurately calibrated. The FLAME process was performed to suspend the input inductors L$_{in}$ and L$_{deg}$ on a single die originally fabricated on a HR-SOI substrate. After FLAME, it can be observed that both |S$_{21}$| and Noise figure improve. The reduction in noise figure is ~0.09 dB making the LNA with suspended inductors comparable to TR-SOI substrate. An improvement of 0.25 dB in |S$_{21}$| is also obtained from S-parameter measurement. It is superior even to TR-SOI substrate by 0.1 dB.

The improvements obtained by substrate removal of L$_{in}$ and L$_{deg}$ are attributed to improvement of Q-factor. The noise figure and gain after substrate removal is simulated using Friis formula as elaborated in Appendix A.1. It can be seen that the measured data agrees well with the simulation. The Q-factor of L$_{in}$ before and after substrate removal is plotted in Fig. 12. It can be seen that, in the LNA operating range, the Q-factor improvement is ~16% and the noise performance and gain is equivalent to TR-SOI substrate. In order to obtain better noise figure, higher Q-factor of L$_{in}$ is necessary. Another noise figure simulation is performed with L$_{in}$ having higher Q-factor and this is also shown in Fig. 11. For this simulation, the inductance value is not changed but the Q-factor measured for inductor 2T-M is used. This is because the inductance value used in present design is nearly same as that of 2T-M with the difference being only 0.3 nH. Hence, this simulation represents a realistic case where Q-factor of L$_{in}$ inductor is optimized by ~81%. With optimized L$_{in}$, the noise figure improvement over HR-SOI and TR-SOI substrate is 0.15 dB and 0.05 dB respectively. The gain only improves marginally by 0.06 dB after L$_{in}$ membrane is replaced by same inductance with Q-factor equivalent to 2T-M. It is to be noted that in this simulation, only L$_{in}$ is optimized while L$_{deg}$ is not changed. By optimizing L$_{deg}$, even better noise figure and gain can be obtained. While the reported improvements are not drastic, it is to be noted that this LNA demonstration is based on an existing design for quick evaluation. The important point to be highlighted is that the performance using FLAME membranes can be better than TR-SOI substrate which is currently the state-of-the-art substrate for RF applications. It is finally enlightening to note that for a very low noise figure, incremental improvement margins require increasingly demanding design and technological approaches [38] [41].

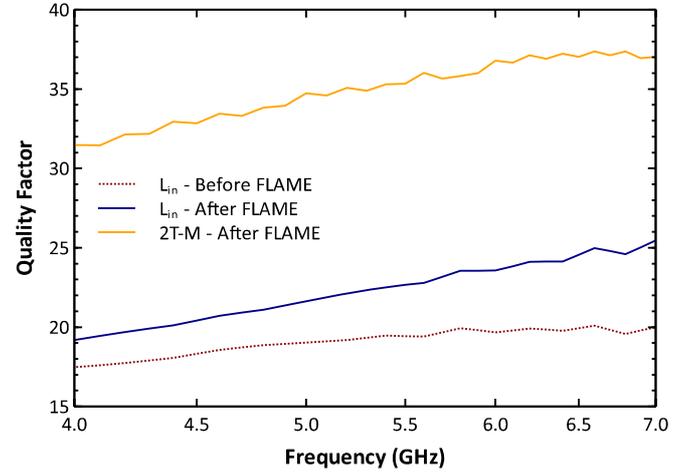

Fig. 12. Measured Q-factor of L$_{in}$ before and after FLAME process plotted along with Q-factor of 2T-M inductor after FLAME

While this study focuses mainly on the noise figure optimization, linearity also has to be considered. The 1-dB compression point (P$_{1dB}$) and output referred third order intercept point (OIP$_3$) has been measured at a frequency of 5.4 GHz. The P$_{1dB}$ and OIP$_3$ values on HR-SOI substrate before FLAME are -5.4 dBm and 9.2 dBm, respectively. After substrate removal P$_{1dB}$ reduces by ~0.3 dB while OIP$_3$ increases by ~0.3 dB. Hence, the linearity of the device is still overall balanced after substrate removal. All these results indicate that amelioration of substrate losses in inductors can be used to augment the LNA performance.

At this stage, it is important to note that the total removal of the handler can potentially cause thermal management problems at high power rating since the silicon whose thermal conductivity is ~150 W/m$^{-1}$K$^{-1}$ is replaced by an interface between the BOX and the air whose heat exchange properties are governed by a conducto-convection mechanism. However, this is not the case in the context of this study, nor for RF switch applications where increased electrical performance in terms of losses and harmonic rejection has been obtained up to an input power of 29 dBm [42].

## VI. CONCLUSION

A substrate removal method has been developed to locally remove handler substrate under inductors to obtain membranes on SOI. A 92% improvement of Q-factor is obtained for a single turn 0.85 nH inductor. For two-turn inductors, the improvement in Q-factor is 35-58% depending on size. A compact model extraction methodology has been developed to obtain lumped element models of the inductors with good correlation to measured data. It has been shown that losses in TR-SOI substrate is not negligible. The application of suspended inductors in LNA circuit has been demonstrated showing that improved Q-factor of the inductors results in better noise performance.

The universal applicability of FLAME process makes it an attractive tool for enhanced RF circuits at high frequencies. The FLAME process also allows to systematically quantify the impact of substrate of each individual component within a given RF circuit. For instance, in this work, different configurations of inductor membranes have been studied to understand the relative importance of each inductor to LNA performance. Hence, FLAME process can also be used as a tool for designers to have direct knowledge of the impact of substrate within various components of the circuit. This is very useful as substrate models are not readily available.

## APPENDIX

*A.1 Noise figure simulation*

The noise figure simulation covered in section V of the manuscript consists of two parts:
1. Simulating noise figure after substrate removal of input inductors ($L_{in}$ and $L_{deg}$) to correlate with measured data
2. Estimating the noise figure improvement with modified $L_{in}$ where inductor membranes with higher Q-factors are incorporated in the LNA design

The methodology for both parts are the same. It is based on the Friis formula for analysis of the noise figure. Measurements of noise figure and S-parameters are used as a reference to compute change in noise figure when Q-factor of input inductor ($L_{in}$) changes. In reality, two inductors form part of the LNA input side ($L_{in}$ and $L_{deg}$) for the inductive source generation topology and both contribute to noise figure [39]. While it is straightforward to compute the impact of Q-factor of $L_{in}$ on noise figure, detailed analysis is necessary when both $L_{in}$ and $L_{deg}$ change. Hence, in our analysis, we keep $L_{deg}$ fixed and analyze the impact of $L_{in}$ on noise figure. The reference data that is used for analysis is noise figure data on LNA on HR substrate with substrate removal performed only under $L_{deg}$ inductor along with different Q-factors of $L_{in}$ inductor.

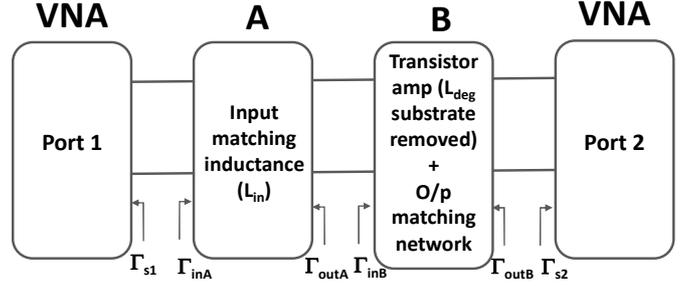

Fig. A.1. Block diagram of LNA for noise figure analysis

The LNA block diagram to compute the noise figure is outlined in Fig. A.1. The different equations that follow are taken from [43]. The expression for noise factor of LNA is given by Friis formula as:

$$F_{LNA} = F_A + \frac{F_B - 1}{G_A} \quad (A.1)$$

$$F_{LNA} = 10^{\frac{NF_{LNA}}{10}} \quad (A.2)$$

$F_A/F_B$ are noise factors of stages A/B and $G_A$ is the power gain of stage A. The LNA noise factor $F_{LNA}$ is calculated from the measured noise figure in dB from equation (A.2). The first step in the analysis is to determine $F_B$ and all other quantities can be deduced by S-parameter measurements made on $L_{in}$ inductor (on HR substrate before substrate removal). The impedance of the inductor is obtained from 1-port measurement on $L_{in}$ inductor as $Z_{ref}$.

From this, the S-parameters of stage A can be computed as:

$$S_{22A} = S_{11A} = \frac{Z_{ref}}{Z_{ref} + 2Z_0} \quad (A.3)$$

$$S_{21A} = S_{12A} = \frac{2Z_0}{Z_{ref} + 2Z_0} \quad (A.4)$$

From these expressions, the noise factors of stage A can be computed as:

$$F_A = \frac{1 - |S_{22A}|^2}{|S_{21A}|^2} \quad (A.5)$$

The input matching network, i.e. the network A, consists of an inductance whose role is to be conjugately matched with the capacitive input impedance of the transistor, i.e. the input impedance of network B. It turns out that $\Gamma_{inB} = \Gamma^*_{outA}$ and hence the power gain of stage A is the available gain $G_{AA}$ given by:

$$G_A = G_{AA} = \frac{|S_{21A}|^2 (1 - |\Gamma_{s1}|^2)}{|1 - S_{11A}\Gamma_{s1}|^2 \cdot (1 - |\Gamma_{outA}|^2)} \quad (A.6)$$

$$\Gamma_{outA} = S_{22A} + \frac{S_{12A} S_{21A} \Gamma_{s1}}{1 - S_{11A}\Gamma_{s1}} \quad (A.7)$$

$\Gamma_{s1} = 0$ because the VNA port impedance is matched to 50 Ω. Under this condition, $\Gamma_{outA} = S_{22A}$ and relation (5) can be further simplified to give:

$$G_A = \frac{|S_{21A}|^2}{(1 - |S_{22A}|^2)} \quad (A.8)$$

Now, equation (A.1) can be modified to obtain $F_B$ as:

$$F_B = (F_{LNA} - F_A)G_A + 1 \quad (A.9)$$

After calculating $F_B$, we can calculate the noise factor of the LNA for any arbitrary values of Q-factor for inductor $L_{in}$.
Both simulations can be performed by calculating impedance of $L_{in}$.

$$Z_{ind} = R_{ind} + jX_{ind} \quad (A.10)$$
$$X_{ind} = imag(Z_{ref}) \quad (A.11)$$
$$R_{ind} = \frac{X_{ind}}{Q_{ind}} \quad (A.12)$$

The corresponding S-parameters are calculated using equations (A.3) and (A.4) with $Z_{ref}$ replaced by $Z_{ind}$. Then, $F_A$ and $G_A$ are calculated using equations (A.5) and (A.8). $F_B$ is already known from reference measurement and does not depend on $L_{in}$. The total noise factor is calculated using equation (A.1).

For simulation type 1, $Z_{ind}$ is obtained by Q-factor measurement with substrate removed for $L_{in}$. For simulation type 2, the Q-factor values for 2T-M with substrate removed is used. This is because the inductance value of 2T-M is very close to $L_{in}$ used in the present design and hence a good representation of Q-factor values that can be practically attained.

The gain can also be simulated within the framework of the noise figure simulation. The total gain of the LNA is given by:
$$G_{LNA} = G_A G_B \quad (A.13)$$
From the S-parameters measurements of the LNA with $L_{deg}$ substrate removed, $G_{LNA}$ is known and from equation (A.8), $G_A$ is also known. From this $G_B$ is calculated. With change in Q-factor of $L_{in}$, the gain $G_A$ is computed using (A.8) with $Z_{ref}$ replaced by $Z_{ind}$. $G_B$ remains the same while the total LNA gain is given by (A.13).

## ACKNOWLEDGEMENTS

This work was supported by: i) the STMicroelectronics-IEMN common laboratory ii) the French government through the National Research Agency (ANR) under program PIA EQUIPEX LEAF ANR-11-EQPX-0025 and iii) the French RENATECH network on micro and nanotechnologies. The authors would like to thank Raphael Paulin from STMicroelectronics for LNA design and concept discussions.